# NEW BULGARIAN-AUSTRIAN PROJECT 'JOINT OBSERVATIONS AND INVESTIGATIONS OF SOLAR CHROMOSPHERIC AND CORONAL ACTIVITY'

Rositsa Miteva[1], Werner Pötzi[2], Astrid Veronig[2], Kamen Kozarev[1], Momchil Dechev[1], Robert Jarolim[2], Mohamed Nedal[1], Nikola Petrov[1], Stefan Purkhard[2], Christoph Schirninger[2], Tsvetan Tsvetkov[1], Yovelina Zinkova[1]

[1]*Institute of Astronomy with National Astronomical Observatory – Bulgarian Academy of Sciences, Sofia, Bulgaria*
[2]*Institute of Physics and Kanzelhöhe Observatory for Solar and Environmental Research, University of Graz, Austria*
*e-mail: rmiteva@nao-rozhen.org*

***Abstract:*** *We present the bilateral collaboration between Bulgarian and Austrian solar and space weather researchers on the topic of chromospheric and coronal activity. This new project will focus, on one hand, on the technical setup and calibration of the new Rozhen chromospheric telescope at the National Astronomical Observatory (NAO) by means of establishing optimal observational programs for different quiet-Sun and activity phenomena, automating the data collection and storage, implementing machine/deep learning models for feature recognition. The second aim is to carry out joint scientific analyses of solar phenomena using observations from ground-based instruments in both countries, and supplementary spacecraft data. The successful implementation of solar monitoring at NAO-Rozhen will facilitate the overall visibility of the Bulgarian instrument and generate interest towards astronomy and solar physics not only for PhD students and young scientists but also for the general public.*

# НОВ БЪЛГАРО-АВСТРИЙСКИ ПРОЕКТ "СЪВМЕСТНИ НАБЛЮДЕНИЯ И ИЗСЛЕДВАНИЯ НА СЛЪНЧЕВАТА ХРОМОСФЕРА И КОРОНАЛНА АКТИВНОСТ"

Росица Митева[1], Вернер Пьотци[2], Астрид Верониг[2], Камен Козарев[1], Момчил Дечев[1], Роберт Яролим[2], Мохамед Недал[1], Никола Петров[1], Щефан Пуркхард[2], Кристоф Ширнингег[2], Цветан Цветков[1], Йовелина Зинкова[1]

[1]*Институт по астрономия с Национална астрономическа обсерватория – Българска академия на науките, София, България*
[2]*Институт по физика с Kanzelhöhe обсерватория, Университет Грац, Австрия*
*e-mail: rmiteva@nao-rozhen.org*

*Ключови думи: Слънчева физика, космическо време*

***Резюме:*** *Представяме двустранното сътрудничество между български и австрийски изследователи на тема хромосферна и коронална активност. От една страна, този нов проект ще се фокусира върху техническото оборудване и калибровка на новия хромосферен телескоп в Националната астрономическа обсерватория (НАО) - Рожен, по-конкретно чрез установяване на оптимални наблюдателни програми за различни събития на спокойното и активно Слънце, автоматизиране на събирането и съхранението на наблюдателните данни, прилагане на модели за машинно обучение при разпознаването на различни характеристики. Втората цел е да се извърши съвместен научен анализ на слънчеви събития чрез наблюдения от наземни инструменти в двете страни, придружено със спътникови данни. Успешното изграждане на програма за слънчеви наблюдения в НАО-Рожен ще подпомогне цялостната видимост на българския телескоп и ще стимулира интерес към астрономията и слънчевата физика не само от докторанти и млади учени, но и от широката публика.*

**Introduction**

We present a bilateral project on the topic of solar and space weather research, supported by the Bulgarian National Science Foundation (BNSF, https://bnsf.bg/) and Austria's Agency for Education and Internationalisation (OeAD, https://oead.at/en/) as part of the Austria-Bulgaria scientific and technological cooperation under a competition procedure that took place in 2022. Science teams from the Institute of Astronomy with NAO - BAS (IANAO) in Bulgaria and the Institute of Physics with Kanzelhöhe observatory for solar and environmental research (KSO), University of Graz in Austria will take part in the project which focuses on ground-based observations of the low corona (chromosphere) complemented with coronal data provided by satellites.

A variety of instruments are used to monitor the Sun, both ground-based (in white light, H-alpha (Hα, 6563 Å) and radio wavelengths) and satellite ((extreme) ultraviolet (E)UV), soft X-rays (SXRs), hard X-rays (HXRs), and gamma-rays). Multiwavelength observations have proven to be the norm for operational and scientific investigation of solar activity. The tasks planned under this joint collaboration include research of quiet (sunspots, quescent filaments, spicules) and eruptive processes on the Sun: solar flares, eruptive filaments, coronal mass ejections (CMEs), in situ solar energetic particles (SEPs), based on their chromospheric and coronal signatures. Ground-based imaging of the Sun is complementary to space-bourne monitoring of solar activity. Its main advantages in the era of space-borne instruments are the relatively low cost of implementation and the ability to access the received data immediately with local computer facilities. Furthermore chromospheric observations are normally not part of space based missions, as they focus mainly on wavelengths that are not observable from ground.

**Aims**

The goals of the proposed project are two-fold:
1. To set up the Rozhen Chromospheric Telescope (RCT), e.g., Petrov (2021), and develop standardized solar observing methodology and products, complementary to the Kanzelhohe Patrol Instrument (KPI), e.g. Pötzi et. al. (2020), by means of strong technical cooperation between the team members. The experience of the Austrian team in solar monitoring and analysis will be invaluable for the timely and successful initiation of the solar observations in Bulgaria (NAO-Rozhen, Fig. 1 left). Training and know-how exchange will be carried out during the entire duration of the project. The new observational data products in NAO-Rozhen will be organized in a complementary way to those in KSO (Fig. 1, right), which will complement and facilitate the combined use of the data products from both observatories. Testing and joint observational campaigns are also planned during the project, in order to provide a proof-of-concept for the selected monitoring programs or optimize them if needed.
2. To carry out combined solar observations with the two instrument suites and external (freely available space-based) resources, in order to study chromospheric signatures of quiet Sun and pre-eruptive active regions (ARs) and multi-wavelength manifestation of solar eruptive phenomena, their morphology and kinematics. These will include studies of the chromospheric magnetic network, solar spicules, the pre-eruptive configurations and dynamics of AR magnetic fields.

To achieve the scientific goal of the proposed project, we will use remote solar observations with high spatial and temporal resolution to characterize the early stages of coronal eruption events in a systematic way by studying the pre-eruptive behavior of filaments and flares during the energy build-up phase, the kinematics and morphology of CMEs and compressive shock waves, and the signatures of high energy non-thermal particles in both remote and in situ observations.

**Project structure and timeline**

The project is organized under 3 work packages (WPs) consisting of different tasks (Ts):
***WP1: Technical support of NAO-Rozhen Chromosphere Telescope and observation campaigns with KSO facilities***
- T1.1: Telescope installation

This task comprises the technical and mechanical part of the project. In order to obtain observations of sufficient quality the new 30cm reflector telescope has to be installed and tested.

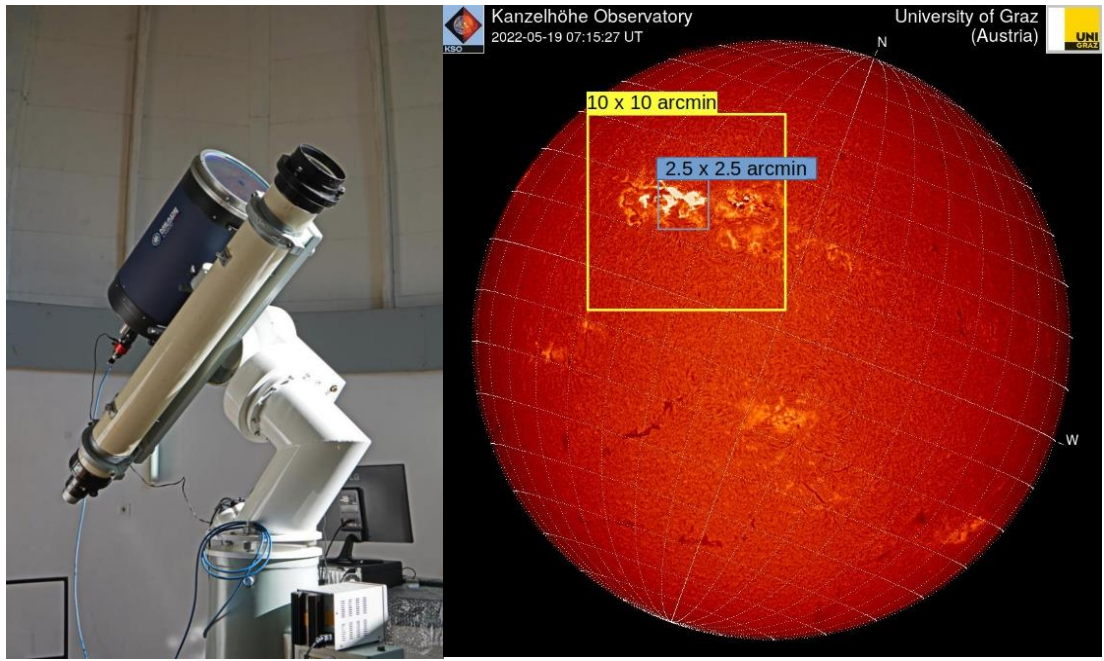

Fig. 1. Left: The 30-cm chromospheric telescope, installed in the solar tower in the National Astronomical Observatory-Rozhen, Bulgaria. Right: A solar observation snapshot made with the Kanzelhöhe Patrol Instrument's Hα telescope showing the full solar disk together with the two fields of view provided by the new Rozhen Chromospheric Telescope, both overlaid on an active region.

- T1.2: Data processing

This task contains the software part of the telescope operation and the data processing. The main goal is the development and documentation of observation modes, data and metadata formats and products, their subsequent (automated) inspection, qualification and storage for the case of NAO-Rozhen solar observations. Databases or catalogs are planned to be developed.

- T1.3: Observation Campaign

This task outlines the observational part of both observatories NAO-Rozhen and KSO. In order to profit from both, the full disk observations at KSO and the high resolution observations at NAO-Rozhen, campaign observations are planned.

- T1.4: Image enhancement

During this task we will develop and apply deep learning algorithms that make use of the short-exposure high frame rate observations (>4/sec) that both KSO and RCT systems can achieve, in order to reconstruct enhanced images from the multi-frame data, with a cadence of a few seconds. The method will be based on the Image-to-Image translation algorithm recently developed by the Graz group (Jarolim et al. 2023).

***WP2 - Joint investigations of solar chromospheric and coronal activity***

- T2.1: Chromospheric Signatures of Quiet Sun and Pre-Eruptive Configurations

The preliminary plan is to perform observations and data analyses of the morphology, dynamics and evolution of pre-flare activity; dynamics of spicules in polar and equatorial zones; sunspot morphology and evolution.

- T2.2: Multi-wavelength study of solar activity phenomena, their morphology and kinematics

This task includes 3 different types of data analyses:

(1) Single event studies - data analyses of case studies of solar flares and filaments using both ground-based and space-based observations, in particular from NASA's Solar Dynamics Observatory and ESA's Solar Orbiter mission. The team members will investigate the energy release and particle acceleration in solar flares, with propagation either towards the Sun as diagnosed in HXRs (e.g. by Solar Orbiter), or escaping the Sun and observed with radio images and spectra (e.g., Low Frequency Array - LOFAR), as well as measured in-situ as SEPs at various locations in the heliosphere.

(2) Analyses of KSO archive data (e.g., sunspots, filaments) available at KSO will also be explored in terms of occurrences, overall properties, association with other solar activity phenomena.

(3) Other statistical studies, e.g., the analyses on white-light (WL) flares, which are an enhancement of the visible continuum in a solar flare, and are mainly associated with large flares, will also be investigated. The emission mechanism of WL flares has not been well understood yet. Some studies have suggested that their origin is accelerated non-thermal electrons and that a relatively strong magnetic field exists in the acceleration site (e.g. Watanabe et al., 2017). During the project we will perform literature review of all reported WL flares starting with the time period of the last solar cycle 24. We will collect and analyze data from available ground-based and satellite databases in order to compile a catalog and perform a statistical study of the observed WL flares.

*WP3: Dissemination of the project results*
- T3.1: Project web-site

In this task, the recently created web-site (https://astro.bas.bg/project-sun/), will be populated with general information, description of activities and results from the project. The maintenance of the web-site will be supported in the future by IANAO.
- T3.2: Scientific dissemination

This task will keep track of all team members' participation in solar/space weather-related conferences, University seminars and other scientific events. Copies and links to the materials will be freely provided via the project web-site.

The project started in mid-August 2023 with a total duration of 2 years. In WP1, T1.1 and T1.2 are scheduled to take part during the first year, whereas T1.3 and T1.4 - during the second year. The activities in the remaining WPs are expected to be carried out during the entire duration of the project.

**Expected impact**

The proposed work is important and timely, as it will contribute significantly to advancing the research on the topic of early-stage solar eruption evolution and energetic particle acceleration. By making the needed connection between the observational parameters of the onset of eruptive filaments and flux ropes, EUV waves/shocks, and SEP acceleration efficiency, the proposed work will enhance our understanding of arguably the most important and consequential stages of solar eruption, which have not been studied in such detail in multi-wavelength analyses previously. This work will be relevant and necessary for interpreting observations from the ongoing NASA and ESA space missions PSP and Solar Orbiter, respectively, that study solar corona and solar wind activities with unprecedented details. Identification of remotely observable coronal eruption parameters that control early particle acceleration will allow us in future work to develop predictive tools that will improve the current state of solar event forecasting. In addition, the project will produce analyzed event samples, which may be used as catalogs in upcoming studies of solar eruptions, and may serve to identify additional predictive parameters of these eruptions in the future.

Furthermore, the joint efforts between the project team members will aid the successful establishment of solar monitoring at NAO-Rozhen and facilitate the overall visibility of the new instrument and the solar group at IANAO. The optical solar observations will also serve as a focal point for the multi-instrument research center for the Sun and space weather planned in IANAO, consisting of a suite of additional instruments: radio antennas (the planned LOFAR station, https://lofar.bg/), neutron detector (https://helio.astro.bas.bg/) and a number of atmospheric monitoring devices.

**Acknowledgements**

The activities under this bilateral cooperation are supported by the Bulgarian National Science Foundation project No. KP-06-Austria/5 (14-08-2023) and Austria's Agency for Education and Internationalisation (OeAD) project No. BG 04/2023.

**References:**

1. Petrov (2021), Sun and Solar Activity: Opportunities for Observations and Development, Publications of the Astronomical Observatory of Belgrade, vol. 100, pp. 137–144.
2. Pötzi *et al.* (2021), Kanzelhöhe Observatory: Instruments, Data Processing and Data Products. *Sol. Phys.* 296, 164 https://doi.org/10.1007/s11207-021-01903-4
3. Jarolim *et al.* (2023) Probing the solar coronal magnetic field with physics-informed neural networks. *Nat. Astron.* https://doi.org/10.1038/s41550-023-02030-9
4. Watanabe *et al.* (2017), Characteristics that Produce White-light Enhancements in Solar Flares Observed by Hinode/SOT, *ApJ* 850, 204 https://doi.org/10.3847/1538-4357/aa9659